\documentclass[reprint,amsmath,amssymb,floatfix,superscriptaddress,notitlepage,showpacs,twocolumn]{revtex4-1}
\usepackage{graphicx}
\usepackage{hyperref}
\usepackage{bm}
\usepackage{amssymb}

\begin{document}

\title{Stability Bounds on Compact Astrophysical Objects from Information-Entropic Measure}

\author{Marcelo Gleiser}
\email{mgleiser@dartmouth.edu}
\affiliation{Department of Physics and Astronomy, Dartmouth College,
Hanover, NH 03755, USA}

\author{Nan Jiang}
\email{Nan.Jiang.GR@dartmouth.edu}
\affiliation{Department of Physics and Astronomy, Dartmouth College,
Hanover, NH 03755, USA}

\date{\today}

\begin{abstract}

We obtain bounds on the stability of various self-gravitating astrophysical objects using a new measure of shape complexity known as configurational entropy. We apply the method to Newtonian polytropes, neutron stars with an Oppenheimer-Volkoff equation of state, and to self-gravitating configurations of complex scalar field (boson stars) with different self-couplings, showing that the critical stability region of these stellar configurations obtained from traditional perturbation methods correlates well with critical points of the configurational entropy with accuracy of a few percent or better. 

\end{abstract}

\pacs{04.40.Dg,11.10.Lm, 03.65.Ge,02.30.Nw}

\maketitle

\section{Introduction}
The issue of gravitational stability, how different assemblies of massive particles and objects can attain a gravitationally-stable state, has been one of the key problems of astrophysical and cosmological research since the late 17th century. Shortly after the publication of his monumental {\it Principia}, Isaac Newton was led to consider the possibility of a spatially infinite universe due to the instability of a finite self-gravitating sphere of matter \cite{Newton}. Barely a year after Einstein published his landmark paper on the general theory of relativity, he examined the gravitational stability of a static, spherically-symmetric universe \cite{Einstein}. That the solution was unstable to perturbations led Einstein to include the so-called cosmological term, which remains a viable explanation to the accelerating recession of far-away Type Ia supernovae \cite{DE1,DE2}, although other explanations based on slowly evolving fields are also consistent with current data \cite{Quintessence}. Moving from the universe to the stability of compact astrophysical objects--the focus of the present manuscript--much depends on the object's specific material composition. In general, the object is modeled with an effective equation of state which attempts to describe the essential physical processes taking place in its interior.

Without presuming to offer here a complete history of how gravitational stability of stellar objects has been examined over the past decades (for reviews see Refs. \cite{Weinberg,Shapiro}), we note that an essential aspect of such stability is that the object's binding energy 
$E_b$ must be negative-definite, $E_b \equiv M - Qm < 0$. (We take $c=\hbar=1$ throughout, unless explicitily shown.) If $E_b>0$ the star is unstable to fission. Here, $Q$ is the conserved number of particles in the object, either baryon number in ordinary stars or the net number of spin-0 bosons in boson stars  \cite{BSreviews}, $M$ is the object's mass, and $m$ is the mass of the particle.

It is also well-known that the negativity of the binding energy is a necessary but not sufficient condition for stability: configurations that have $E_b < 0$ may still be unstable under radial perturbations. In order to establish the stability of stellar configurations one needs to apply perturbations to the effective equations describing the self-gravitating matter. Leaving details aside (the interested reader can consult Refs. \cite{Weinberg,Shapiro}), the key point is that unstable stellar configurations will have exponentially-growing radial perturbations characterized by imaginary eigenvalues of the perturbed linearized equations. Stability conditions are established imposing that the perturbations conserve particle number, as in Chandrasekhar's variational formalism \cite{ChandraVR}. 

In the present work we examine the stability of self-gravitating objects from a very different perspective. Instead of the usual perturbative approach, we will apply a recently-proposed measure of shape complexity known as configurational entropy (CE) \cite{GS1} to stellar-like objects, investigating the stability of both Newtonian and fully relativistic objects. In particular, we will expand the results of Ref. \cite{GSo1} and apply our formalism to three classes of objects: general Newtonian polytropes that model non-relativistic and ultra-relativistic white dwarfs \cite{Weinberg,Shapiro}; neutron stars modeled by an Oppenheimer-Volkoff equation of state \cite{OV}; and to boson stars, self-gravitating configurations made of charge-conserving complex scalar fields \cite{Kaup,Ruffini,Colpi,Gleiser,GleiserWatkins,KusmartsevMielke}.
 
We will show that the configurational entropy, an extension of Shannon's information entropy \cite{Shannon} to spatially-localized mathematical functions based on their Fourier transforms, can provide reliable bounds on the stability of self-gravitating objects with accuracies of a few percent or better. As such, the information-entropic method used here provides an alternative approach to the study of gravitational stability with broad applicability.

This paper is organized as follows. In Section II we briefly review the formalism  and obtain the equations describing general relativistic, sphericaly-symmetric compact objects. In Section III we review the notion of configurational entropy (CE). In Section IV we apply the formalism to cold white-dwarfs, showing how the CE can be used to obtain an estimate of Chandrasekhar's critical stability mass for these objects \cite{GSo1}. We show that the mass for polytropes scales inversely with their CE, allowing us to relate the mass instability region -- a saddle ridge in configuration space -- to an equivalent instability region in the object's configurational entropy. In Section V we investigate neutron stars modeled with an Oppenheimer-Volkoff (OV) equation of state, showing how the CE gives a bound on the compact object's stability consistent with the traditional perturbation method. We also show that an inverse scaling relation similar to that found for Newtonian polytropes relating the object's mass and its CE holds for OV neutron stars. In Section VI we apply the formalism to boson stars with self-coupled scalar fields. We show how the CE again gives reliable bounds on the compact object's critical stability mass and how the same scaling between mass and CE found for polytropes and OV neutron stars is also applicable for these objects. We conclude in Section VII with final remarks and a discussion of future projects.

\section{General Formalism}

We consider static, spherically-symmetric configurations with spacetime metric (we follow the conventions of ref. \cite{Weinberg}),
\begin{equation}
\label{Metric}
ds^2=B(r)dt^2-A(r)dr^2-r^2(d\theta^2+sin^2\theta d\phi^2),
\end{equation}
\noindent
and take $c=\hbar=1$. Einstein's field equations are
\begin{equation} 
\label{EinsteinEquations}
G_{\mu\nu}=-8\pi GT_{\mu\nu}, 
\end{equation}
\noindent
where $T_{\mu\nu}$ is the energy-momentum tensor. For Newtonian polytropes and neutron stars, we will model stellar matter as a perfect fluid with energy-momentum tensor
\begin{equation}
\label{Energy-Momentum}
T_{\mu\nu} = p(r)g_{\mu\nu} + \left [p(r) + \rho(r)\right ]U_{\mu}U_{\nu},
\end{equation}
\noindent
where $p(r)$ is the pressure, $\rho(r)$ is the energy density and $U^{\mu}$ is the velocity four-vector. Taking the star to be at rest,  $U^{\mu}$ has only one non-zero component, $U_0= - \sqrt{B(r)}$. For boson stars, the energy-momentum tensor is computed from a Lagrangian density to be defined later. We use the energy density to define the mass of the object as
\begin{equation}
\label{Mass}
M = 4\pi\int_0^R\rho(r)r^2dr~,
\end{equation}
\noindent
where the upper limit of integration, $R$, is to either the object's radius $R$, where $\rho(R)=0$ or, for boson stars, to $R = \infty$, although most of the star's mass is concentrated within an effective radius $R_{\rm eff}\equiv \int_0^{\infty}\rho(r)r^3dr/\int_0^{\infty}\rho(r)r^2dr$.

With these definitions, Einstein's equations can be written as:
\begin{equation}
\label{StellarEquations}
\begin{split}
&\frac{1}{A}(\frac{A^\prime}{Ar}-\frac{1}{r^2})+\frac{1}{r^2}=8\pi G\rho;\\
&\frac{1}{A}(\frac{B^\prime}{Br}+\frac{1}{r^2})-\frac{1}{r^2}=8\pi Gp;\\
&\frac{B^\prime}{B}=-\frac{2p^\prime}{p+\rho},
\end{split}
\end{equation}
\noindent
where a prime denotes derivative with respect to the radial direction. The last expression is the equation for hydrostatic equilibrium. These equations, together with an equation of state $p(r) = p[\rho(r)]$, are used to study a large variety of self-gravitating objects, assuming that $A(0)=1$ and $B(r\rightarrow\infty)=1$. The equation involving $A(r)$ and $\rho(r)$ may be integrated as $A(r) = [1-2G{\cal M}(r)/r]^{-1}$, where the mass density function is given by ${\cal M}(r) \equiv \int_0^r4\pi r'^2\rho(r')dr'$. 

As is well-known, Eqs. \ref{StellarEquations} can be rearranged and, using the above expression for $A(r)$, the gravitational fields $A(r)$ and $B(r)$ can be eliminated to obtain \cite{Weinberg}
\begin{eqnarray}
\label{Hydrostatic}
-r^2p'(r)&=&G{\cal M}(r)\rho(r)\left [1+\frac{p(r)}{\rho(r)}\right ]\nonumber \\
&&\left [1+\frac{4\pi r^3p(r)}{{\cal M}(r)} \right ] \left [1-\frac{2G{\cal M}(r)}{r}
\right ].
\end{eqnarray}

This equation describes self-gravitating stellar configurations with general-relativistic corrections in the last three terms. We are interested here only in isentropic stars, that is, those with a constant entropy per particle across the star. Such configurations model very low temperature white dwarfs and neutron stars, as well as boson stars, which are self-gravitating spin-0 boson-condensates. Next we review the main ideas behind the configurational entropy measure of spatial complexity before we use it to establish stability bounds for all three types of configurations.\\

\section{Configurational Entropy}
Since we are interested in self-gravitating configurations with spatially-localized energy, consider the set of square-integrable bounded functions $f({\bf x}) \in L^2({\bf R})$ and their Fourier transforms $F({\bf k})$. Plancherel's theorem states that
\begin{equation}
\int_{-\infty}^{\infty}|f({\bf x})|^2d^dx = \int_{-\infty}^{\infty}|F({\bf k})|^2 d^dk.
\end{equation}
Now define the modal fraction $f({\bf k})$ \cite{GS1},
\begin{equation}
\label{modal fraction}
f({\bf k}) = \frac{|F({\bf k})|^2}{\int|F({\bf k})|^2 d^dk},
\end{equation}
where the integration is over all ${\bf k}$ where $F({\bf k})$ is well-defined and $d$ is the number of spatial dimensions. $f({\bf k})$ measures the relative weight of a given mode ${\bf k}$. This can also be seen by noting that $|F({\bf k})|^2$ is proportional to the Fourier transform of the two-point correlation function of the function $f({\bf x})$, while $\int_{-\infty}^{\infty}|F({\bf k})|^2 d^dk$ is the integrated power \cite{GSo2}.
For periodic functions where a Fourier series is defined, $f(k)\rightarrow f_n=|A_n|^2/\sum |A_n|^2$, where $A_n$ is the coefficient of the $n$-th Fourier mode.

We define the configurational entropy $S_C[f]$ as \cite{GS1}
\begin{equation}
\label{entropy for discrete modes}
S_C[f] = - \sum f_n \ln (f_n),
\end{equation}
in analogy with Shannon's information entropy, $S_S = - \sum p_i \log_2 p_i$ \cite{Shannon}. Note that if all $N$ modes $k$ carry the same weight $f_n = 1/N$, the discrete configurational entropy has a maximum at $S_C=\ln N$. If only one mode is present, $S_C = 0$. These limits motivate the definition of Eq. \ref{entropy for discrete modes}.

For general, non-periodic functions in the continuous interval $(a,b)$, the case of interest here, the configurational entropy $S_C[f]$
is \cite{GS1}
\begin{equation}
\label{configurational entropy}
S_C[f] = - \int {\tilde f}({\bf k})\ln [{\tilde f}({\bf k})] d^d k,
\end{equation}
where ${\tilde f}({\bf k})=f({\bf k})/f({\bf k})_{\rm max}$ and $f({\bf k})_{\rm max}$ is the maximum fraction, in many cases of interest given by the zero mode, ${\bf k}=0$, or by the system's longest physical mode, $|k_{\rm min}| =\pi/R$. This normalization guarantees that ${\tilde f}({\bf k})\leq 1$ for all physical values of ${\bf k}$.
We call $\sigma({\bf k}) = -{\tilde f}({\bf k})\ln [{\tilde f}({\bf k})]$ the configurational entropy density. In this paper, we will compute the configuration entropy from the energy density $\rho(r)$ of the self-gravitating object. The choice of the energy density is the most natural, given that it is a spatially-localized function that encapsulates all the relevant physics and boundary conditions describing the stellar configuration. The task at hand is thus to solve the relevant Einstein's equations to obtain the equilibrium configurations in terms of $\rho(r)$ and then use $\rho(r)$ to compute the CE as a function of the star's central density, $\rho(r=0)\equiv \rho_0$. We start with the simplest case, Newtonian polytropes modeling cold white dwarfs.

\section{Cold White Dwarfs and the Chandrasekhar Limit}

Newtonian polytropes are obtained from the hydrostatic equation (setting the general relativistic corrections to zero in Eq. \ref{Hydrostatic}) \cite{Weinberg,Shapiro}
\begin{equation}
\frac{d}{dr}\left[\frac{r^2}{\rho(r)}\frac{dp(r)}{dr}\right]=-4\pi Gr^2\rho(r).
\label{hydroeq}
\end{equation}
Eq. \ref{hydroeq} is supplemented by a general polytropic equation of state
\begin{equation}
p=K\rho^{\gamma},
\label{polyeq}
\end{equation}
where the constant $K$ depends on the entropy per nucleon and chemical composition. No heat flow throughout the object requires  $\gamma$ to be the adiabatic index, defined as the ratio of the heat capacities of the fluid at constant pressure and volume. Small mass, stable non-relativistic  white dwarfs are well-modeled by $\gamma=5/3$ and $K=\frac{\hbar^2}{15m_e\pi^2}\left(\frac{3\pi^2}{m_N\mu}\right)^{5/3}$, where $m_{e(N)}$ is the electron (nucleon) mass, and $\mu\sim 2$ is the number of nucleons per electron. The largest mass white dwarfs are modeled by $\gamma=4/3$ and  $K= \frac{\hbar}{12\pi^2}\left(\frac{3\pi^2}{m_N\mu}\right)^{4/3}$, the well-known Chandrasekhar limit \cite{Weinberg,Shapiro}. The binding energy for polytropes with $Q$ nucleons, $E_b=M-Qm_N$, can be written as $E_b=-\frac{(3\gamma-4)}{(5\gamma-6)}\frac{GM^2}{R}$, where $M$ is given by Eq. \ref{Mass}. There is a clear stability boundary at $\gamma=4/3$ where $E_b$ changes sign. We will show below that the configuration entropy captures the same boundary.

\begin{figure}[htbp]
\includegraphics[width=\linewidth]{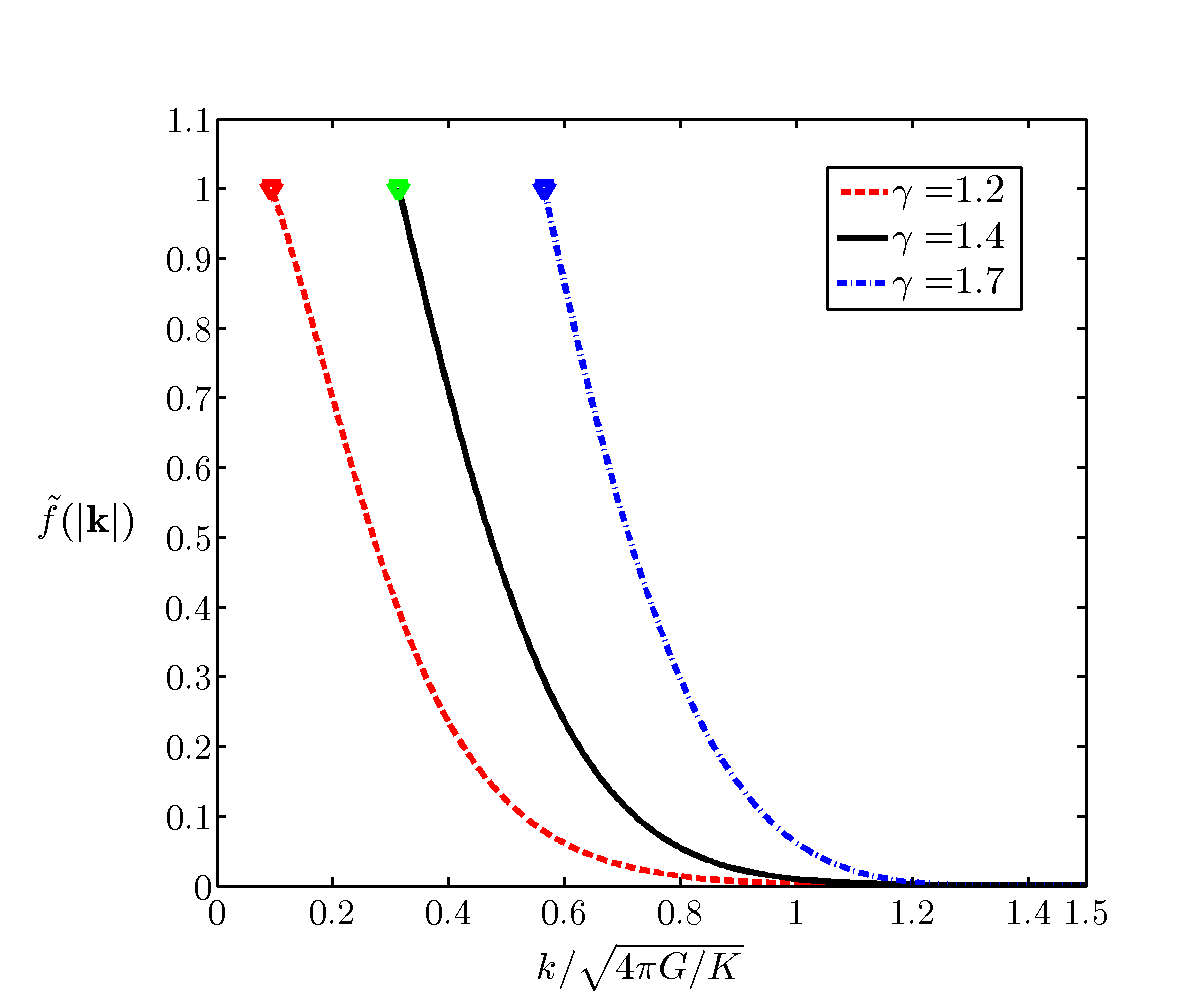}
\caption{(Color online.) Normalized modal fraction ${\tilde f}(|{\bf k}|)$ for sample values of the polytropic index $\gamma$. From left to right, $\gamma=1.2, 1.4, {\rm and} ~1.7$.}
\label{FTPoly}
\end{figure}

Solutions to Eqs. \ref{hydroeq} and \ref{polyeq} must satisfy $\rho(0)=\rho_0$ and $\rho'(0)=0$, and are found introducing new variables $\rho(r) = \rho_0\theta(\xi)^{1/(\gamma-1)}$ and $\xi=r/\alpha$, with $\alpha^2=\frac{K\gamma}{4\pi G(\gamma-1)}\rho_0^{(\gamma-2)}$. Equation \ref{hydroeq} then becomes the Lane-Emden equation with boundary conditions $\theta(0)=1$ and $\theta'(0)=0$,
\begin{equation}
\frac{1}{\xi^2}\frac{d}{d\xi}\xi^2\frac{d\theta}{d\xi} + \theta^{1/(\gamma-1)}=0.
\label{Lane-Emden}
\end{equation}
Solutions were obtained via a 4$^{th}$-order Runge-Kutta method with step size $10^{-3}$. The CE is computed from the energy density using Eq. \ref{configurational entropy}. Since polytropes have a finite radius (where $\rho(R)=0$ or, equivalently, $\theta(\xi_R)=0$, with $\xi_R\equiv R/\alpha$), the $k$-integration is in the interval $k \in [k_{\rm min}=\pi/R,\infty)$. This ensures that only modes with wavelengths smaller than the polytrope contribute to the configurational entropy \cite{GSo1}. In Fig. \ref{FTPoly} we plot the normalized modal fraction ${\tilde f}(|{\bf k}|)$ for sample values of the polytropic index $\gamma$. The infrared  cutoff is at $k_{\rm min} =\pi/R$, defined by the star's radius. The configurational entropy for the various polytropes is computed using this modal fraction to integrate the CE density as described in section III. 

We next present a simple scaling argument relating the stellar mass to its configurational entropy. Using the dimensionless variables $\theta(\xi)$ and $\xi$ in Eq. \ref{Mass} we obtain, 

\begin{equation}
\begin{split}
M &= 4\pi\int_0^{R}\rho(r)r^2dr\\
& = 4\pi\rho_0\alpha^3\int_0^{\xi_R}\theta^{1/(\gamma-1)}(\xi)\xi^2d\xi\\
&\propto \rho_0\alpha^{3}\propto \rho_0^{(3\gamma-4)/2}.
\label{Mscaling}
\end{split}
\end{equation}
Using the dimensionless variables in the Fourier transform of the energy density, we can express the modal fraction as
\begin{equation}
\begin{split}
{\tilde f}(k) &= \frac{h(\alpha k)}{h(\frac{\alpha\pi}{ R})}= \frac{h(\alpha k)}{h(\frac{\pi}{\xi_R})}\\
& = \frac{h(\alpha k)}{C(\gamma)},
\end{split}
\end{equation}
where $C(\gamma)$ is independent of $\rho_0$, and $h(\alpha k)$ is
\begin{equation}
h(\alpha k) = \left|\int_0^{\xi_R}\theta^{1/(\gamma-1)}(\xi)\exp({i\alpha k\cdot\xi})\xi^2d\xi\right|^{2}.
\end{equation}
\noindent
The configurational entropy is then,
\begin{equation}
\begin{split}
S &= -4\pi\int_{k_{min}}^{\infty} \frac{h(\alpha k)}{C(\gamma)}\log\left(\frac{h(\alpha k)}{C(\gamma)}\right)k^2dk\\ 
&= -4\pi\alpha^{-3}\int_{\kappa_{min}}^{\infty} \frac{h(\kappa)}{C(\gamma)}\log\left(\frac{h(\kappa)}{C(\gamma)}\right)\kappa^2d\kappa\\
\end{split}
\end{equation}
where $\kappa = \alpha k$, so that $\kappa_{min} = \pi/\xi_R$. We thus obtain,
\begin{equation}
S\rho_0^{-1} \propto\alpha^{-3}\rho_0^{-1}\propto\rho_0^{(4-3\gamma)/2}.
\end{equation}
Comparing with Eq. \ref{Mscaling}, we see that $S\rho_0^{-1} \propto M^{-1}$. Note that at $\gamma = 4/3$ the quantity $S\rho_0^{-1}$ is independent of $\rho_0$, consistent with a boundary in the star's stability \cite{Weinberg}: as is well-known, stars with $\gamma < 4/3$ are unstable, while stars with $\gamma > 4/3$ are stable. $\gamma=4/3$ defines an instability ridge for the family of stellar configurations, as we show next by exploring how both the mass and the configurational entropy vary with central density and $\gamma$.

\begin{figure}[htbp]
\includegraphics[width=\linewidth]{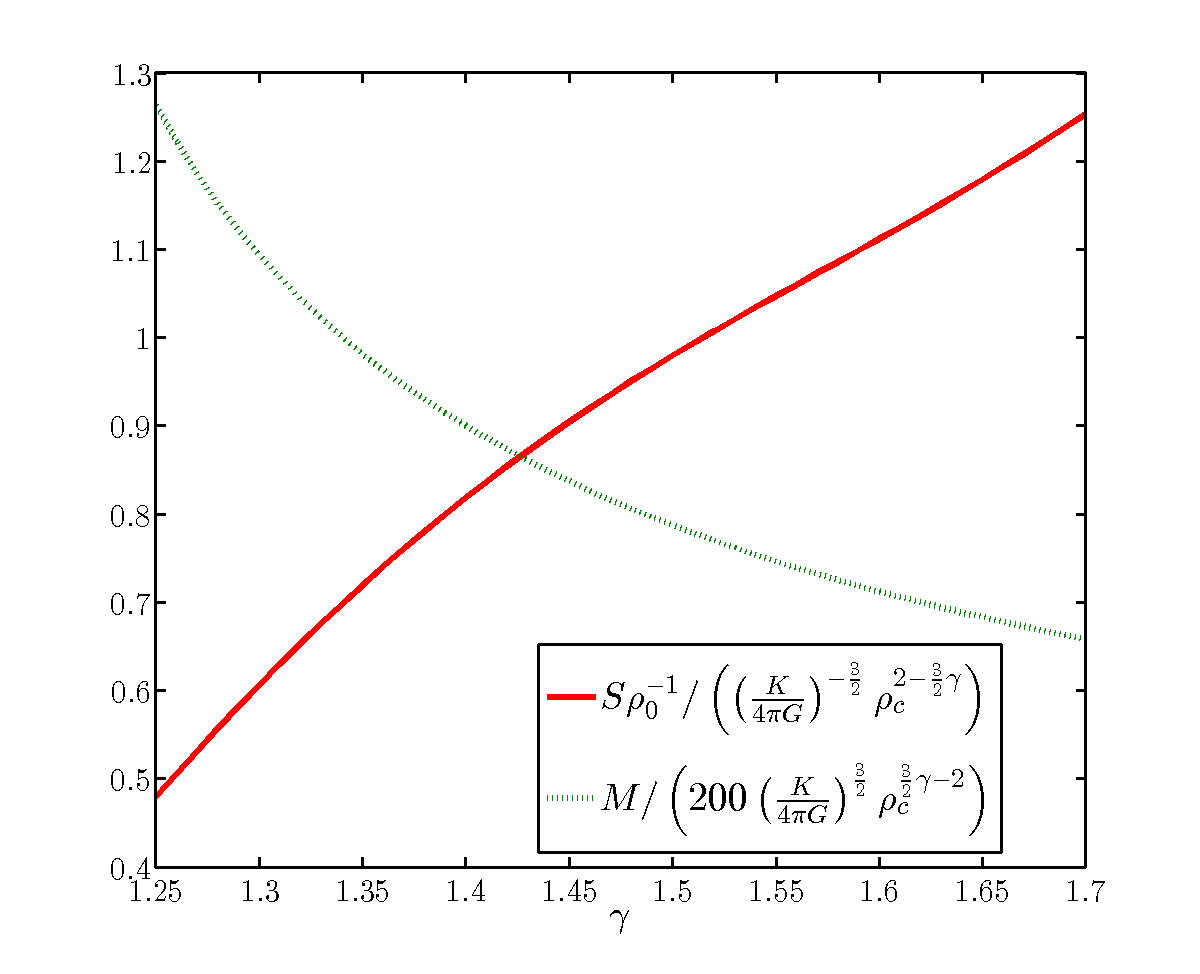}
\caption{(Color online.) Configurational entropy times $\rho_0^{-1}$ (continuous line) and mass (dotted line) versus polytropic index $\gamma$ for $\rho_0=\rho_c$.}
\label{CE_MassPoly}
\end{figure}

The mass and configurational entropy $S\rho_0^{-1}$ are shown as a function of polytropic index $\gamma$ in Fig. \ref{CE_MassPoly}, with $\rho_0=\rho_c$, where $\rho_c$ is a fiducial value for the critical central density, which can be computed for a few specific values of $\gamma$. For example, for $\gamma=5/3$, $\rho_c=0.97\times 10^6\mu$ gm/cm$^3$, where $\mu\simeq 2$ is the number of nucleons per electron \cite{Weinberg}. 

\begin{figure}[htbp]
\includegraphics[width=\linewidth]{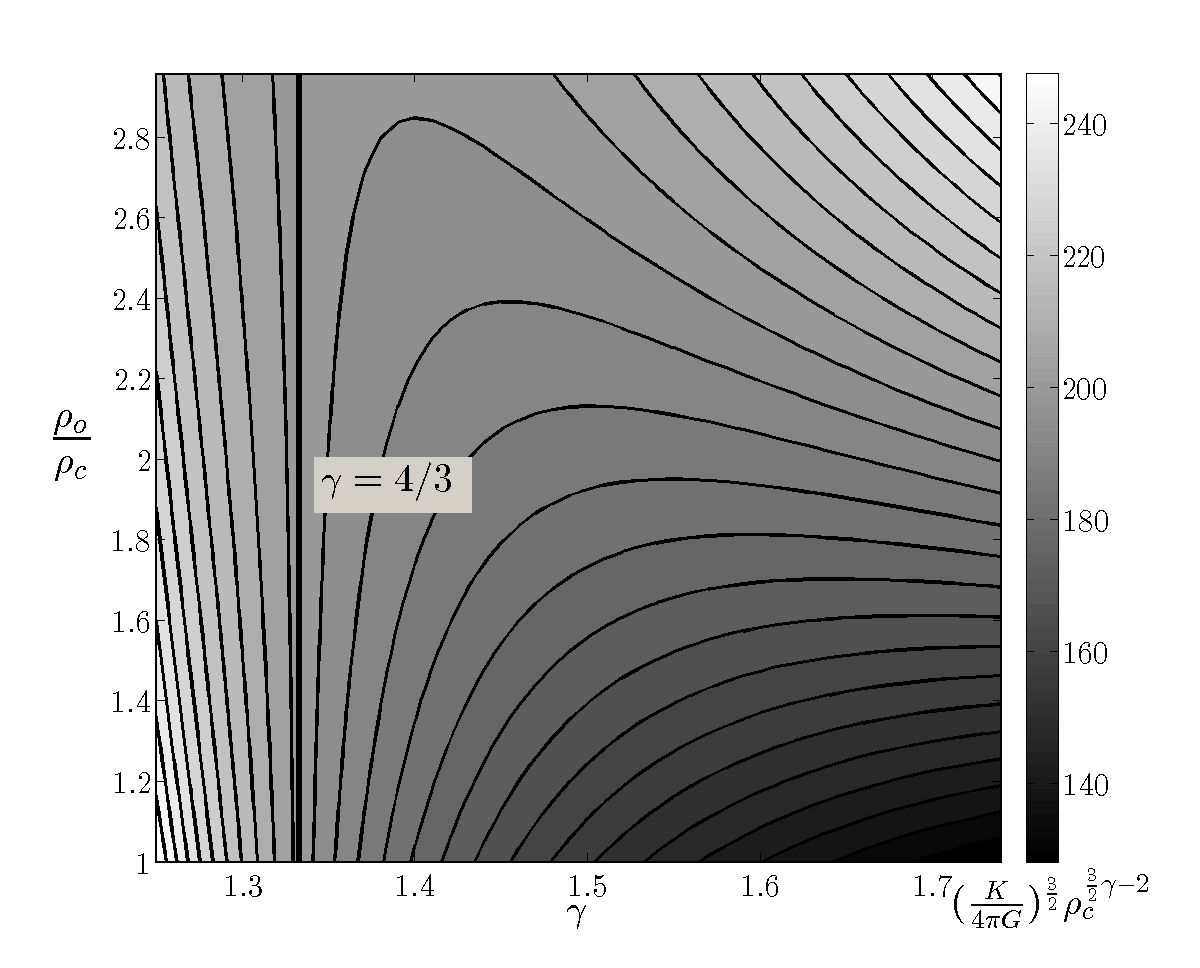}
\caption{Contour plot for the mass of polytropes as a function of the central density $\rho_0/\rho_c$ and polytropic index $\gamma$. There is an instability ridge--a saddle line--for $\gamma=4/3$. The vertical bar specifies the values in units given at its bottom.}
\label{M_Poly}
\end{figure}

\begin{figure}[htbp]
\includegraphics[width=\linewidth]{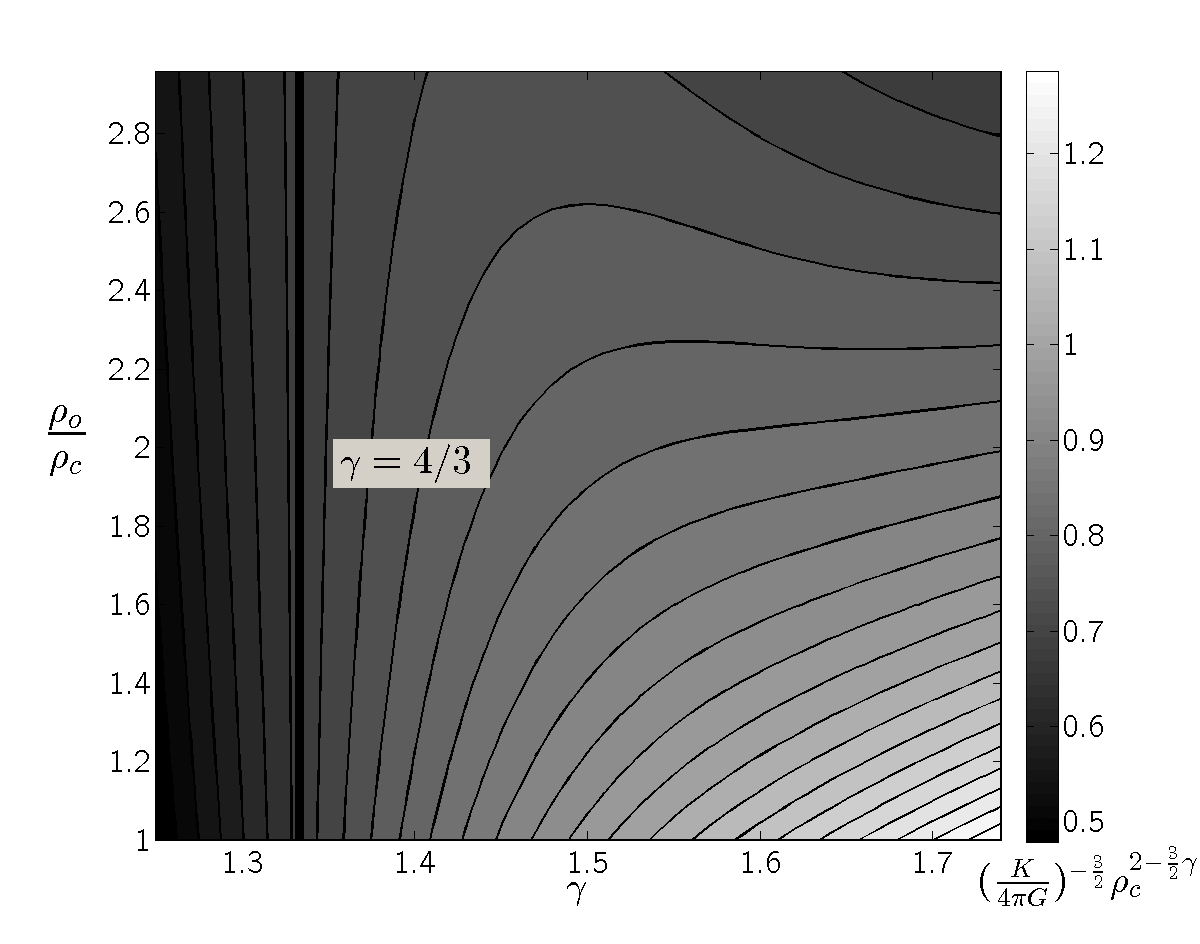}
\caption{Contour plot for the configurational entropy of polytropes as a function of the central density $\rho_0/\rho_c$ and polytropic index $\gamma$. There is an instability ridge--a saddle line--for $\gamma=4/3$. The vertical bar specifies the values in units given at its bottom.}
\label{CE_Poly}
\end{figure}

In Fig. \ref{M_Poly} we show the contour plot of the stellar mass as a function of $\rho_0/\rho_c$ and $\gamma$, where the existence of a saddle ridge at $\gamma=4/3$ is clear. In Fig. \ref{CE_Poly} we show the contour plot of the quantity $S\rho_0^{-1}$ as a function of $\rho_0/\rho_c$ and $\gamma$. The reader can verify that the shadings are approximately reversed for the two plots (there are small deviations due to the $gamma$-dependence of the relevant quantitities), illustrating qualitatively the inverse scaling between mass and configurational entropy discussed above.

\begin{figure}[htbp]
\includegraphics[width=\linewidth]{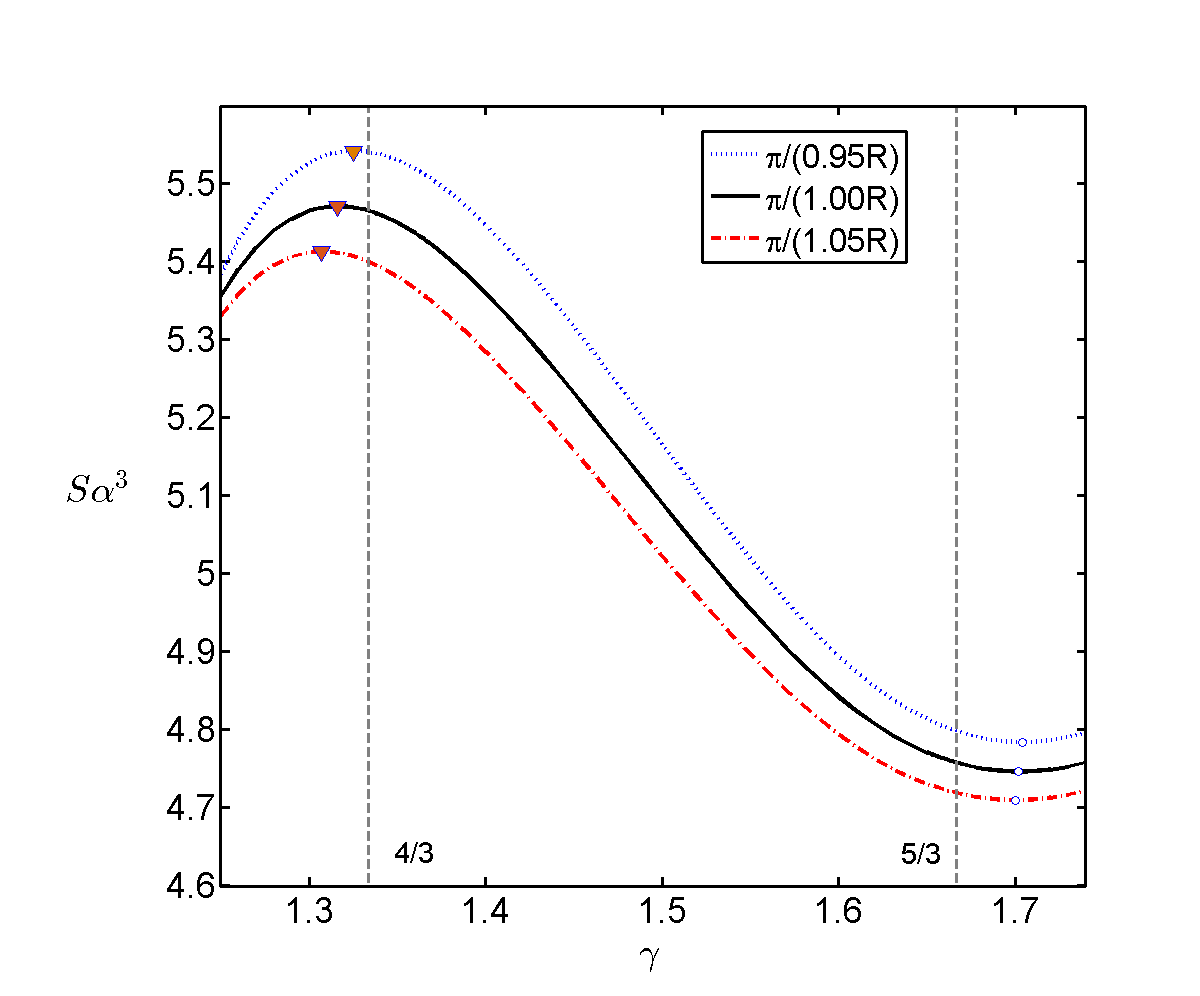}
\caption{(Color online.) Configurational entropy versus polytropic index $\gamma$ for polytropes. We display results for several choices of cutoff for $k_{\rm min}$.}
\label{CEgamma}
\end{figure}

In Fig. \ref{CEgamma} we plot the configurational entropy versus polytropic index $\gamma$ for various choices of cutoff for $k_{\rm min}$. We do this to illustrate the sensitivity of the results to the choice of cutoff and to establish that it is best to choose what is physically more natural, that is, no arbitrary fine-tuning of cutoff, and thus $k_{\rm min} = \pi/R$. Even if the results are not perfectly accurate with this choice, they lie within a few percent from the critical points for CE: the maximum of CE lies at $1.3\%$ from $\gamma=4/3$ -- the polytropic index for an ultra-relativistic white dwarf -- and the minimum of CE at $2.1\%$ from $\gamma=5/3$, the polytropic index for the most stable nonrelativistic white dwarf. In particular, the CE has a maximum for $\gamma = 1.316$. Since the mass decreases with $\gamma$ for fixed $\rho_0/\rho_c$ (cf. Fig. \ref{CE_MassPoly}), the CE gives an upper bound on the maximum mass for stability, $M_{\rm max}$. This mass approximates the Chandrasekhar mass ($M_{\rm Chandra}$) at $\gamma=4/3$ as $M_{\rm max} = 1.0373 M_{\rm Chandra}$, so within $3.73\%$ of the correct value. We also correct a typo in Ref. \cite{GSo1} in the vertical axis, which should read $S\alpha^3$ and not $SR^3$. 

We see that the CE accurately captures the stability properties of Newtonian polytropes. This gives us confidence that we can use similar methods to examine the stability of general relativistic compact objects, which we do next.

\section{Neutron Stars with Oppenheimer-Volkoff Equation of State}

As we move into general-relativistic objects, we start with a simple but representative model, neutron stars with an Oppenheimer-Volkoff (OV) equation of state, where neutrons are treated as a pure ideal gas \cite{OV}. Much work has been done in the past decades extending the results of OV to more realistic situations, where calculations try to incorporate a variety of effects taking into account the role of the strong nuclear force at the star's core. These approaches are treated in many books and reviews, such as those listed in references \cite{Shapiro, Lattimer}. Our interest at this point is not to explore different equations of state modeling a neutron star interior, but how the effects of general relativity, in particular their impact on a star's stability, are reflected in its equivalent configurational entropy. Can we obtain information about a compact relativistic object's stability from its information-entropic complexity?

Considering a gas of particles with rest mass $\mu_0$ obeying Fermi-Dirac statistics, the related equation of state may be written in parametric form as \cite{OV}
\begin{equation}
\begin{split}
\label{OVparam}
\rho &=K(\sinh t-t)\\
p &=\frac{1}{3}K(\sinh t-8\sinh \frac{1}{2}t+3t),\\
\end{split}
\end{equation}
where $K=\pi\mu_0^{4}c^5/4h^2$ and 
\begin{equation}
t=4\log \left (\frac{k_F}{\mu_0c} + \left [1 + \left (\frac{k_F}{\mu_0c}\right )^2\right ]^{1/2}\right ),
\end{equation}
with $k_F$ being the maximum momentum in the Fermi distribution, related to the particle number density $n$ as $n = k_F^3/3\pi^2\hbar^3$.
We follow OV and introduce a new mass function variable $u(r)$ (equivalent to the mass density function $G{\cal M}(r)$ defined in Section II),
\begin{equation}
\frac{1}{A}=1-2u/r,
\end{equation}
so that $u$ obeys
\begin{equation}
du/dr=4\pi \rho r^2.
\end{equation}
Einstein's equations (Eqs. \ref{StellarEquations}) become:
\begin{equation}
\begin{split}
\frac{du}{dr} &=r^2(\sinh t-t)\\
\frac{dt}{dr} &=-\frac{4}{r(r-2u)}\frac{\sinh t-2\sinh \frac{1}{2}t}{\cosh t-4\cosh \frac{1}{2}t+3}\\
&\times\left[\frac{1}{3}r^3(\sinh t-8\sinh \frac{1}{2}t+3t)+u\right],\\
\end{split}
\end{equation}
where $K=1/4\pi$, and the units of length $a$ and mass $b$ have been fixed as
\begin{equation}
a=\frac{1}{\pi}\left(\frac{h}{\mu_{0}c}\right)^{3/2}\frac{c}{\left(\mu_{0}G\right)^{1/2}};~
b=\frac{c^2}{G}a.
\end{equation}

\begin{figure}[htbp]
\includegraphics[width=\linewidth]{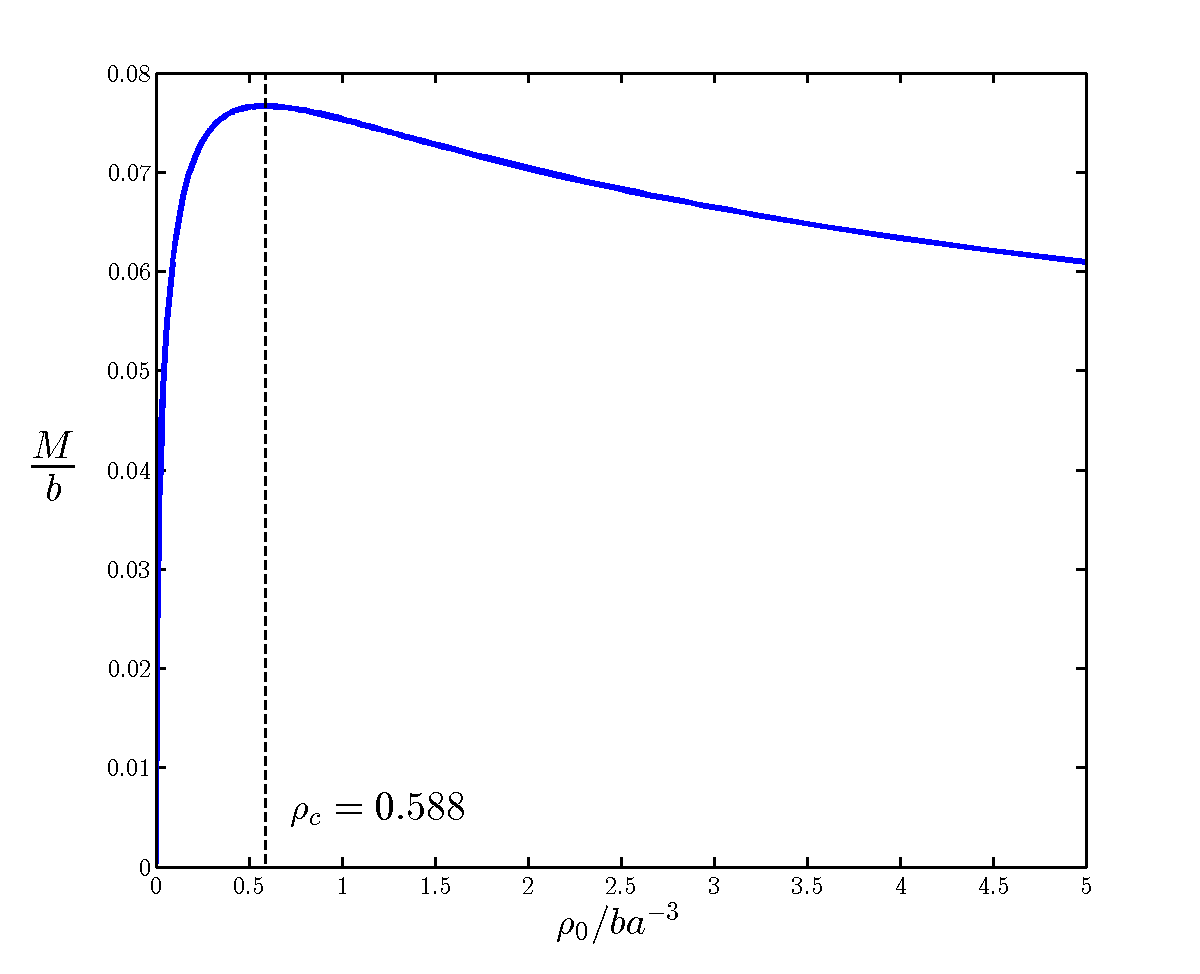}
\caption{OV neutron star mass vs. central density $\rho_0$. Stars with $\rho_0>\rho_c=0.588 ba^{-3}$ are known to be unstable to gravitational collapse.}
\label{NSMass}
\end{figure}

We can now solve numerically Einstein's equations with boundary conditions 
\begin{equation}
\begin{split}
u(r=0)&=0;\quad t(r=0)=t_0\\
u(r=r_b)&=u_b;\quad t(r=r_b)=0,
\end{split}
\end{equation}
so that $p(r=r_b)=\rho(r=r_b)=0$ and $r_b$ is the radius of the star and $u_b$ is its mass. Results are thus parameterized in terms of $t_0$, related to the star's central density $\rho_0$ by Eq. \ref{OVparam}. In Fig. \ref{NSMass} we plot the mass of the OV neutron star for values of the central density parameter $\rho_0$. Stars with $\rho_0>\rho_c$ are perturbatively unstable to gravitational collapse, as is well-known.

\begin{figure}[htbp]
\includegraphics[width=\linewidth]{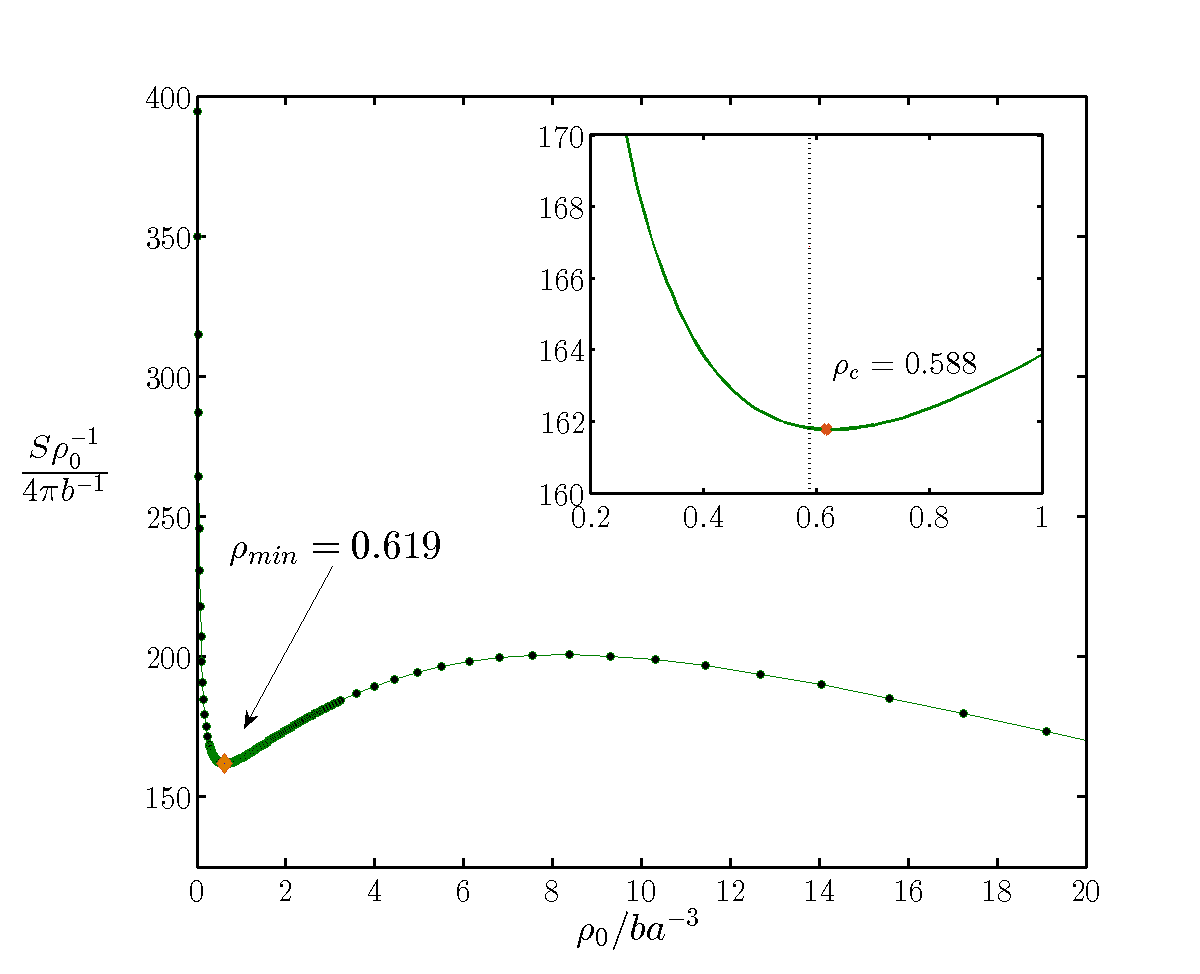}
\caption{Configurational entropy times $\rho_0^{-1}$ for the OV neutron star vs. central density $\rho_0$. Stars with $\rho_0>\rho_c=0.588ba^{-3}$ are perturbatively unstable to gravitational collapse. The inset shows the result near the CE minimum.}
\label{NSCE}
\end{figure}

As with the Newtonian polytropes, we compute the configurational entropy using the energy density of the equilibrium configurations. The range of integration is again $k_{\rm min} = \pi/R \leq k < \infty$, reflecting the fact that neutron stars have well-defined radii where $\rho(R)=0$. The results are shown in Fig. \ref{NSCE}, where we can see that the quantity $S\rho_0^{-1}$ has a minimum ($\rho_{\rm min}$) near the critical equilibrium value of the central density ($\rho_c$) where the stellar mass is a maximum.  The configurational entropy is multiplied by the inverse central density to have a quantity that scales with dimensions of inverse mass.The inset shows the results in more detail near the CE minimum at $\rho_{\rm min}=0.619ba^{-3}$ and thus within 5.3\% from $\rho_c$. We can translate the value of $\rho_{\rm min}$ to an estimate of the critical mass based on CE, thus establishing a bound in the critical OV neutron star mass with accuracy of $0.58\%$. (The flatness near the mass function maximum helps.) 

\begin{figure}[htbp]
\includegraphics[width=\linewidth]{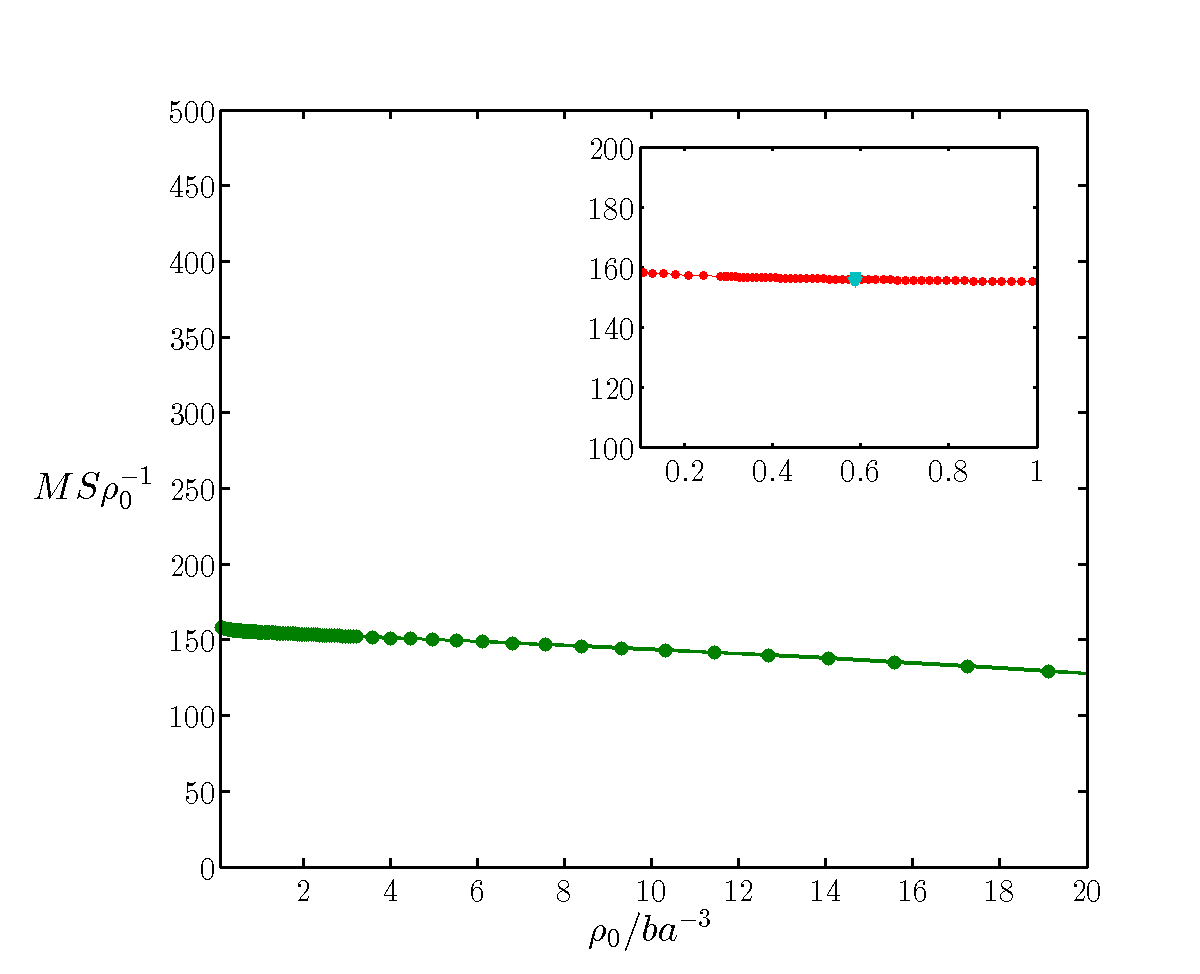}
\caption{Behavior of the quantity $S\rho_0^{-1}M$ as a function of central density $\rho_0$ for OV neutron stars. The approximate linear scaling persists for a wide range of central densities The inset shows the result near $\rho_c=0.588ba^{-3}$, the critical value for stability.}
\label{NS_scaling}
\end{figure}

The reader should not confuse the results of Section IV for Newtonian polytropes, where the prediction for the Chandrasekhar mass was given at the maximum of the CE with respect to the polytropic index $\gamma$, with the results here, where the estimate for the critical mass comes at the minimum of the CE with respect to central density $\rho_0$. The OV equation of state is only well-modeled by a polytrope in the non-relativistic limit for neutrons, with $\gamma=5/3$. For this value of $\gamma$, the star's mass is a decreasing monotonic function of $\rho_0$ \cite{Weinberg}. The maximum mass in the mass vs. central density plot appears only when general relativistic effects are included. From Fig. \ref{NSCE}, that the configurational entropy offers a reliable measure for the stability of OV neutron stars, providing an accurate estimate for the critical mass. We also verified numerically that neutron stars obey the same approximate scaling as Newtonian polytropes, in that the product $S\rho_0^{-1}M$ is nearly constant with $\rho_0$. The result is shown in Fig. \ref{NS_scaling}. This inverse scaling justifies why the critical mass, being a maximum, correlates well with the minimum of the configurational entropy.

We now extend our approach to another class of general relativistic bound objects, boson stars.

\section{Boson Stars}

Boson stars are self-gravitating spheres of scalar fields \cite{Kaup, Ruffini, BSreviews}. These hypothetical objects are possible both in the Newtonian and general-relativistic limits, and for free and self-interacting fields. Due to their remarkable properties, boson stars have attracted much interest over the past decades \cite{BSreviews}. Being made from self-gravitating spin-0 bosons, these objects are supported against gravitational collapse from Heisenberg's uncertainty pressure and, when applicable, from repulsive interactions among the particles \cite{Colpi}. Indeed, for a free complex scalar field with $U(1)$-conserving charge $Q$, these objects have masses $M\sim (M_{\rm Pl}^2/m)$ and radii $R\sim 1/m$, where $m$ is the mass of the particle excitation of the field, and $M_{\rm Pl}$ is the Planck mass \cite{Kaup, Ruffini, BSreviews}. If a repulsive self-interaction is added, the mass scales as $M\sim \lambda^{1/2}(M_{\rm Pl}^3/m^2)$ \cite{Colpi}. Furthermore, boson stars have many qualitative similarities with neutron stars, with a maximum mass $M$ marking the stability boundary against radial perturbations \cite{Gleiser, GleiserWatkins}. Stars made of real scalar fields are also possible, but the configurations are time-dependent, known as oscillatons \cite{Seidel}. Given the many similarities, and the potential applicability of boson stars to many questions of current interest, from being dark matter candidates to serving as exploratory tools probing the boundary between classical and quantum field theory, we will now investigate whether the configurational entropy of boson stars can furnish information about their stability.

\subsection{Formalism}

For completeness, we briefly review the essential formalism to find boson stars. Consider the action
\begin{equation}
S=\int d^4x\sqrt{-g}\{\frac{R}{16\pi G}+\mathcal{L}\},
\end{equation}
where ${\mathcal L}$ is the Lagrangian density, 
\begin{equation}
\mathcal{L}=g^{\mu\nu}\partial_{\mu}\phi\partial_{\nu}\phi^*-m^2|\phi|^2-\frac{\lambda}{4}|\phi|^4.
\end{equation}
We write the spherically-symmetric complex scalar field as $\phi(r,t)=\Phi(r)e^{-i\omega t}$, where $\Phi(r)$ is real and has no nodes. This means that we will only be investigating here the stability properties of boson stars in their ground state. Stars can be found in excited states and their decay properties have interesting consequences, including the generation of gravitational wave bursts \cite{FerrellGleiser}. We define the dimensionless variables $x =mr$ and ${\tilde t} = mt$. Primes are derivatives with respect to $x$. We also absorb the dimensionless frequency ${\tilde \omega} \equiv \omega/m$ into the metric coefficient $B$, ${\tilde B} = B/{\tilde \omega}^2$ and define the dimensionless field $\sigma(x) \equiv \Phi(x)/\sqrt{8\pi G}$. (Henceforth we suppress tildes.) It proves convenient to rewrite the dimensionless coupling constant $\lambda$ as \cite{Colpi}
\begin{equation}
\Lambda = \lambda\frac{M_{\rm Pl}^2}{8\pi m^2}.
\label{Lambda}\end{equation}
With these definitions, variation with respect to the metric of Eq. \ref{Metric} and with respect to the scalar field gives Einstein's equations (Eqs. \ref{EinsteinEquations}) and the Klein-Gordon equation as
\begin{equation}
\begin{aligned}
A^{\prime}&=xA^2\left [\frac{\sigma^{\prime 2}}{A}+(\frac{1}{B}+1)\sigma^2 +\frac{\Lambda\sigma^4}{4}\right ] -\frac{A}{x}(A-1)\\
B^{\prime}&=xAB\left [\frac{\sigma^{\prime 2}}{A}+(\frac{1}{B}-1)\sigma^2 -\frac{\Lambda\sigma^4}{4}\right ] +\frac{B}{x}(A-1)\\
\sigma^{\prime\prime}&=-\left [ \frac{2}{x}+\frac{1}{2}\left (\frac{B^{\prime}}{B}-\frac{A^{\prime}}{A}\right )\right ]\sigma^{\prime}-
A\left [\left (\frac{1}{B}-1\right )\sigma - \frac{\Lambda}{2}\sigma^3\right ].
\label{BSEquations}
\end{aligned}
\end{equation}
These equations are solved for the boundary conditions
$A(0)=1;~B(\infty)=1;~\sigma(0)=\sigma_0;~\sigma(\infty)=0;~\sigma^{\prime}(0)=0$. Note that solutions are parametrized by the central value of the scalar field $\sigma_0$, which determines the star's central density.

\begin{figure}[htbp]
\includegraphics[width=\linewidth]{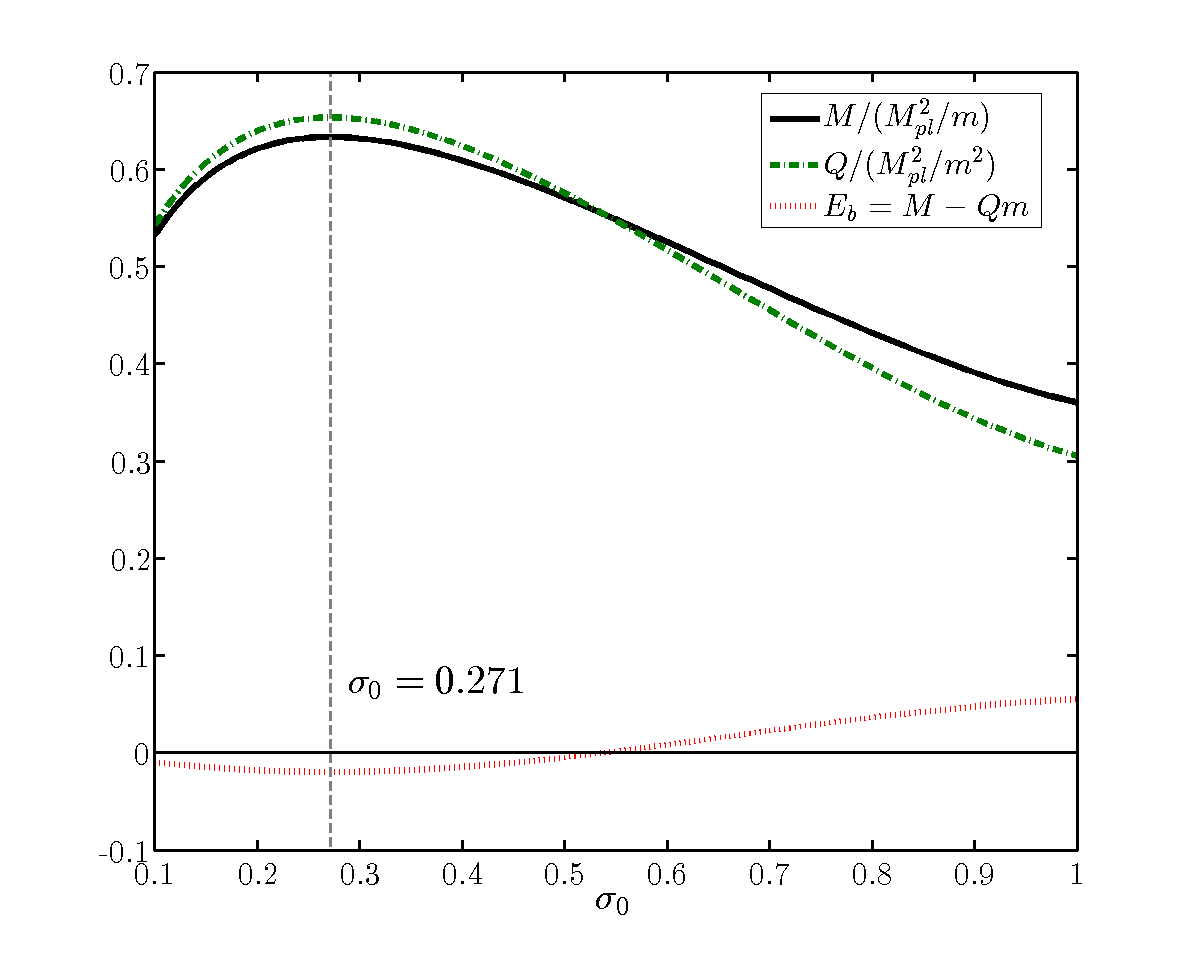}
\caption{Boson star mass (continuous line) and conserved charge $Q$ (dot-dashed line) vs. central value of the scalar field $\sigma_0$. We also show the binding energy $E_b$ (dotted line). Stars with $\sigma_0>\sigma_c=0.271$ are known to be unstable to gravitational collapse.}
\label{BSmassQ}
\end{figure}

In Figure \ref{BSmassQ} we plot the boson star mass and conserved charge for the free field case ($\Lambda =0$). Note that the maximum mass is also where the binding energy $E_b$ is maximal, where $E_b = M -Qm$ (lower line). As shown in Refs. \cite{Gleiser, GleiserWatkins} the maximum mass is also the stability boundary for the boson star. This is also the case for the interacting case, $\Lambda \neq 0$ \cite{Gleiser, GleiserWatkins}. Note also that stars with $\sigma_0>0.540$ have $E_b>0$, and are thus unstable to fission.

\subsection{Configurational Entropy for Boson Stars}

\begin{figure}[htbp]
\includegraphics[width=\linewidth]{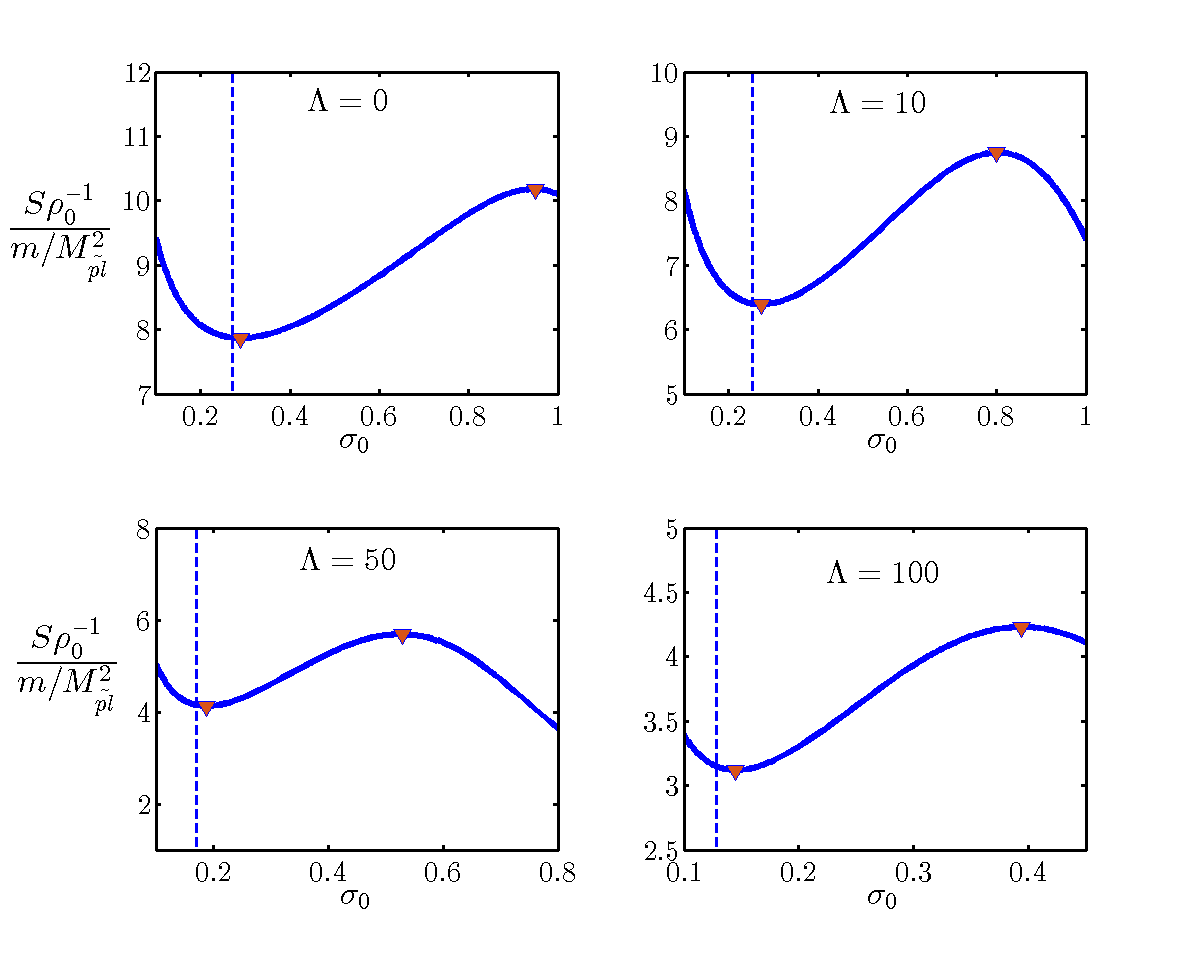}
\caption{The configurational entropy for boson stars multiplied by inverse central density $S\rho_0^{-1}$ as a function of the scalar field's central value $\sigma_0$ for different values of the scalar field coupling $\Lambda$.  The dashed vertical line denotes $\sigma_c$, the instability boundary for the star under radial perturbations. As in the case with neutron stars, the CE provides a reliable estimate for the critical mass, with precision better than $\sim 1\%$. Values of maximum stellar masses for different values of $\Lambda$ are listed in Table \ref{Table}.}
\label{BSCE}
\end{figure}

We now compute the configurational entropy for boson stars from Eq. \ref{configurational entropy} using the Fourier transform of the energy density as we did with Newtonian polytropes and neutron stars. This means that for each value of $\sigma_0$ we find the solution of the coupled Einstein-Klein-Gordon system of equations and use it to compute the star's energy density $\rho(r)$. We do this for several values of the scalar self-coupling. Note that since the scalar field only vanishes at spatial infinity, boson stars don't have a specific radius where the energy density and pressure vanish.  We thus don't use a momentum cutoff, computing the CE for all momenta $0 \leq |{\bf k}| \leq \infty$. The results are shown in Figure \ref{BSCE} as a function of the field's central value $\sigma_0$ for different values of the coupling $\Lambda$. The vertical lines denote the critical value of the field ($\sigma_c$) beyond which the star is unstable under radial, charge-conserving perturbations. It is apparent that these lines are very near the minima of CE for all values of $\Lambda$. Just as for OV neutron stars, the configurational entropy provides a reliable bound on the star's stability.

Specifically, we can use the minimum of the configurational entropy to obtain a bound on the star's maximum mass, as was done previously for neutron stars. The results are summarized in the Table  below for different values of $\Lambda$. The third column is the value of the critical mass obtained from the CE, while the last column gives the percentual error of the estimate.

\vspace{0.5cm}

\hspace{0.5 in}
\begin{table}[htdp]
\begin{center}
\begin{tabular*}{0.25\textwidth}{@{\extracolsep{\fill}} |c|c|c|c|} \hline
$\Lambda$ & $M_{\rm crit}$ & $M_{\rm crit}^{CE}$ & $\Delta (\%)$ \\ \hline
	0	  &	0.6330	    &		0.6324	      &		0.10	       \\ \hline
	10	  &    0.7863	    &	        0.7845	      &	        0.25	       \\ \hline
	50	  &    1.2450	    &          1.2367	      &        0.66	       \\ \hline
	100	  &    1.6522	    &	        1.6351	      &        1.04         \\ \hline
\end{tabular*}
\end{center}
\caption{Comparison of maximum stable mass for boson stars obtained from traditional perturbation methods (second column, $M_{\rm crit}$) and from the configurational entropy (third column, $M_{\rm crit}^{CE}$) for different values of the scalar field self-coupling $\Lambda$. The fourth column list the percentual error of the estimate using the CE.}
\label{Table}
\end{table}

\vspace{0.5cm}

\begin{figure}[htbp]
\includegraphics[width=\linewidth]{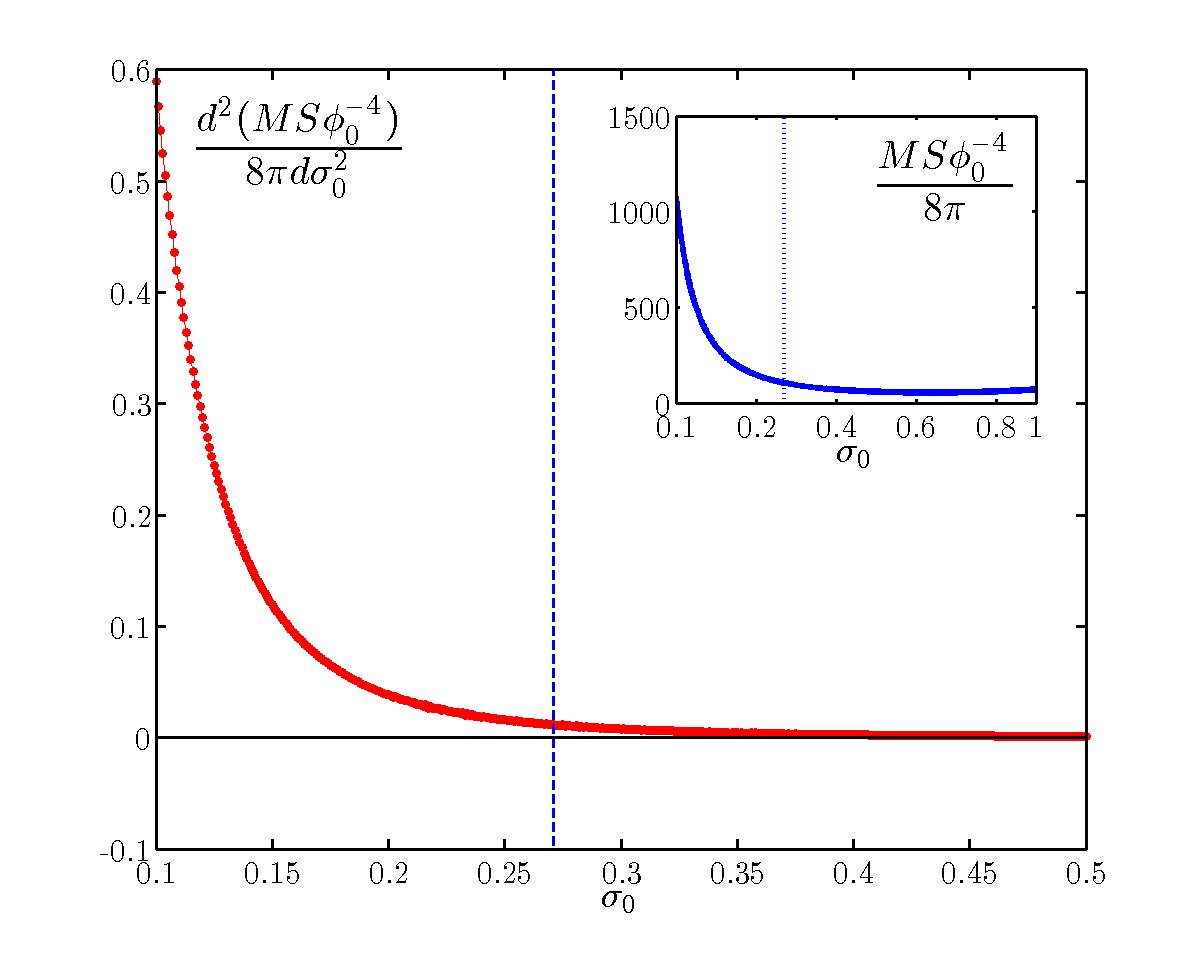}
\caption{The inset shows the quantity $S\phi_0^{-4}M$ as a function of the scalar field's central value $\sigma_0$ for $\Lambda=0$.  The main plot shows the second derivative of $S\phi_0^{-4}M$ with respect to $\sigma_0$, displaying the same saddle ridge behavior found for polytropes.}
\label{BS_scaling}
\end{figure}

\begin{figure}[htbp]
\includegraphics[width=\linewidth]{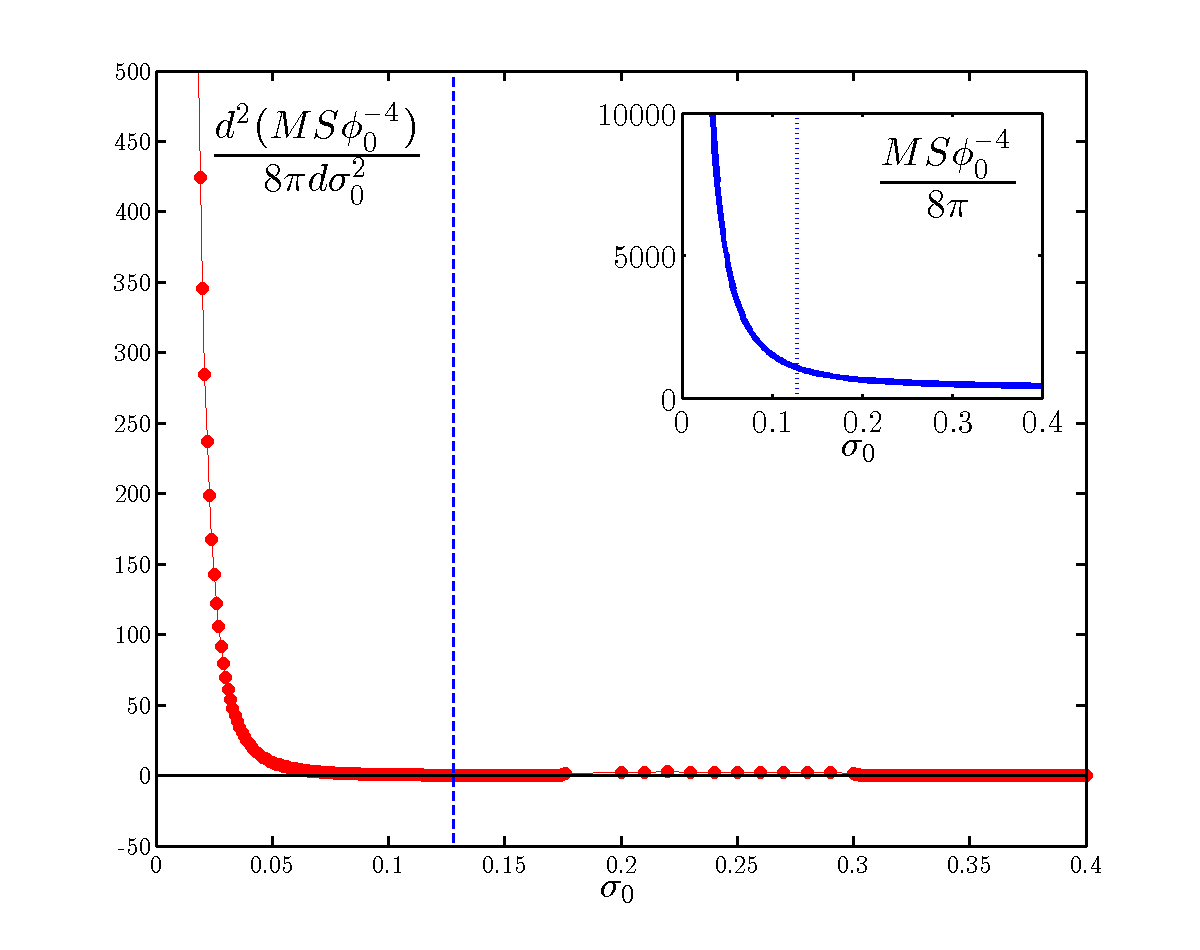}
\caption{The inset shows the quantity $S\phi_0^{-4}M$ as a function of the scalar field's central value $\sigma_0$ for $\Lambda=100$.  The main plot shows the second derivative of $S\phi_0^{-4}M$ with respect to $\sigma_0$, displaying the same saddle ridge behavior found for polytropes.}
\label{BS_Lscaling}
\end{figure}

Given the qualitatively similar results between neutron stars and boson stars, we should be able to show that the quantity $S\sigma_0^{-4}M$ is approximately constant with respect to the central field value $\sigma_0$, leading to the inverse scaling $S\sigma_0^{-4} \sim M^{-1}$ . The fourth power of the field ensures  the same dimensionality as when the energy density is used, as was the case with polytropes and OV neutron stars. The result is shown in the inset of Fig. \ref{BS_scaling}, where it can be seen that an approximate linear scaling holds in the region near and above the critical energy density. The main plot shows the second derivative of the quantity approaching zero near the critical value $\sigma_c=0.271$. In Fig. \ref{BS_Lscaling} we present the equivalent results for stars with $\Lambda = 100$, where a similar scaling holds.

\section{Concluding Remarks}

We investigated the stability properties of a variety of self-gravitating compact objects using the recently-proposed configurational entropy \cite{GS1}, a quantity that computes the relative weights of different Fourier modes making up a given configuration inspired by Shannon's information entropy \cite{Shannon}. We extended previous results for Newtonian polytropes, where the Chandrasekhar mass for ultra-relativistic white dwarfs is estimated to within 3.73\% of the correct value, to fully general-relativistic neutron stars modeled with an Oppenheimer-Volkoff equation of state and to boson stars made of self-interacting complex scalar fields. Using the energy density of the configurations to compute their respective configurational entropy, we were able to obtain predictions to the critical stable mass with precision better than one percent for all these objects. We have further shown that an inverse scaling relation holds between the star's configurational entropy and its mass near the critical region and beyond. This scaling helps clarify why a critical value for the mass is reflected in a critical value for the configurational entropy, although we are still pursuing a first-principles derivation relating the two quantities.

We are current investigating two related questions. First, we are computing the configurational entropy of excited states of boson stars \cite{FerrellGleiser} in order to relate their decay and gravitational radiation emission to their configurational-entropic properties. Preliminary results indicate that the configurational entropy grows with the quantum numbers labeling excited states ($n, \ell, m$), as is the case with simple quantum mechanical systems. It will be interesting to see whether the configurational entropy will provide information to resolve energy-degenerate states, i.e., states with the same binding energy and different quantum numbers. We are also investigating the evolution of the configurational entropy during gravitational collapse. We expect that as the star becomes more localized its configurational entropy will increase. An important question is to determine whether the configurational entropy reaches a maximum when the event horizon forms and whether there is a relation between this hypothetical maximum value and Bekenstein's entropy based on the surface area of the black hole \cite{Bekenstein}.

\acknowledgements MG was supported in part by a Department of Energy grant DE-SC0010386. MG and NJ also
acknowledge support from the John Templeton Foundation grant no. 48038.


\begin{thebibliography}{99}

\bibitem{Newton} I. Newton, {\it Four Letters to Richard Bentley}, in ``Newton'', selected and edited by I. Bernard Cohen and R. S. Westfall (WW Norton, NY 1995).

\bibitem{Einstein} A. Einstein, {\it Cosmological Considerations on the General Theory of Relativity}, in ``The Principle of Relativity: A collection of original papers on the special and general theory of relativity'' (Dover, NY 1952).

\bibitem{DE1} A. G. Riess et al. [Supernova Search Team Collaboration], Astron. J. {\bf 116}, 1009 (1998); S. Perlmutter et al. [Supernova Cosmology Project Collaboration], Astrophys. J. {\bf 517}, 565 (1999). 

\bibitem{DE2} For a textbook treatment of dark energy see, e.g., L. Amendola and S. Tsujikawa, {\it Dark Energy: Theory and Observations} (Cambridge University Press, Cambridge, UK, 2015).

\bibitem{Quintessence} R.R. Caldwell, R. Dave, P.J. Steinhardt, Phys. Rev. Lett. {\bf 80}, 1582 (1998). For recent results from 2015 Planck satellite  see, arXiv:1502.01590v1.

\bibitem{Weinberg} S. Weinberg, {\it Gravitation and Cosmology: Principles and Applications of the General Theory of Relativity} (John Wiley \& Sons, New York, NY, 1972).

\bibitem{Shapiro} S. L. Shapiro and S. A. Teukolsky, {\it Black Holes, White Dwarfs, and Neutron Stars: The Physics of Compact Objects} (John Wiley \& Sons, New York, NY, 1983).

\bibitem{BSreviews} For reviews see, T.D. Lee, Y. Pang, Phys. Rep. {\bf 221}, 251 (1992); A. R. Liddle and
M. S. Madsen, Int. J. Mod. Phys. D1, 101 (1992); P. Jetzer, Phys. Rep. {\bf 220}, 163 (1992). 

\bibitem{ChandraVR} S. Chandrasekhar, Phys. Rev. Lett. {\bf 12}, 437 (1964); Astrophys. J. {\bf 140}, 417 (1964).

\bibitem{GS1} M. Gleiser and N. Stamatopoulos, Phys. Lett. B {\bf 713}, 304 (2012).

\bibitem{GSo1} M. Gleiser, D. Sowinski, Phys. Lett. B {\bf 727}, 272 (2013).

\bibitem{OV} J. R. Oppenheimer and G. M. Volkoff, Phys. Rev. {\bf 55}, 374 (1939).


\bibitem{Kaup} D.J. Kaup, Phys. Rev. {\bf 172}, 1331 (1968).

\bibitem{Ruffini} R. Ruffini, S. Bonazzola, Phys. Rev. {\bf 187}, 1767 (1969).

\bibitem{Colpi} M. Colpi, S.L. Shapiro, I. Wasserman, Phys. Rev. Lett. {\bf 57}, 2485 (1986).

\bibitem{Gleiser} M. Gleiser, Phys. Rev. D {\bf 38}, 2376 (1988).

\bibitem{GleiserWatkins} M. Gleiser, R. Watkins, Nucl. Phys. B {\bf 319}, 733 (1989).

\bibitem{KusmartsevMielke} F. V. Kusmartsev, E. W. Mielke, and F. E. Schunck, Phys. Rev. D {\bf 43}, 3895 (1991).

\bibitem{Shannon} C.E.Shannon, Bell Syst. Tech. J. {\bf 27} (1948) 379; {\it ibid.}, 623.

\bibitem{GSo2} M. Gleiser, D. Sowinski, Phys. Lett. B {\bf 747}, 125 (2015).

\bibitem{Lattimer} J. Lattimer, Annu. Rev. Nucl. Part. Sci. {\bf 62}, 485 (2012).

\bibitem{Seidel} E. Seidel, W.M. Suen, Phys. Rev. D {\bf 42}, 384 (1990); L.A. Ure\~na-L\'opez, Class. Quantum Grav. {\bf 19}, 2617
(2002).

\bibitem{FerrellGleiser} R. Ferrell and M. Gleiser, Phys. Rev. D {\bf 40}, 2524 (1989)

\bibitem{Bekenstein} J. D. Bekenstein, Phys. Rev. D {\bf 7}, 2333 (1973); 

\end{thebibliography}
\end{document}